\documentclass[prd,aps,showpacs,groupedaddress,eqsecnum,notitlepage,nofootinbib]{revtex4-2}
\usepackage{graphicx,bm}
\usepackage[utf8]{inputenc}
\usepackage{enumerate, color}
\usepackage{tikz,siunitx,mwe}
\usepackage{graphicx}
\usepackage{tensor}
\usepackage{epstopdf}
\usepackage{amsmath}
\usepackage{amsfonts}
\usepackage{slashed}
\usepackage{amssymb}
\usepackage{subfigure}
\usepackage{lipsum}
\usepackage{color}
\usepackage{eurosym}
\usepackage{scalerel}%To smaller subindex
\usepackage{float}
\usepackage{accents}
\usepackage[colorlinks=true,pdfstartview=FitV,linkcolor=blue,citecolor=blue,urlcolor=blue,breaklinks=true]{hyperref}
\usepackage{mathtools}%http://ctan.org/pkg/mathtools
\newcommand{\be}{\begin{equation}}
\newcommand{\ee}{\end{equation}}
\newcommand{\ben}{\begin{eqnarray}}
\newcommand{\een}{\end{eqnarray}}
\newcommand{\bes}{\begin{subequations}}
\newcommand{\ees}{\end{subequations}}

\begin{document}
%%%%%%%%%%%%%%%%%%%%%%%%%%%%%%%%%%%%%%%%%%%%%%%
%%%%%%%%%%%%%%%%%%%%%%%%%%%%%%%%%%%%%%%%%%%%%%%
%---------------------------------------------%
\title{Localized scalar structures around static black holes}
%---------------------------------------------%
\author{D. C. Moreira$^{1}$}
\author{F. A. Brito$^{1,2}$}
\author{D. Bazeia$^{2}$}
%---------------------------------------------% 
\affiliation{$^1$Departamento de F\'isica, Universidade Federal da Campina Grande, 58429-900, Campina Grande, PB, Brazil}
\affiliation{$^2$Departamento de F\'isica, Universidade Federal da Para\'iba, 58051-970, Jo\~ao Pessoa, PB, Brazil}
%---------------------------------------------%
\email{moreira.dancesar@gmail.com; fabrito2007@gmail.com; dbazeia@gmail.com}
%%%%%%%%%%%%%%%%%%%%%%%%%%%%%%%%%%%%%%%%%%%%%%%
%%%%%%%%%%%%%%%%%%%%%%%%%%%%%%%%%%%%%%%%%%%%%%%
\begin{abstract}
%%%%%%%%%%%%%%%%%%%%%%%%%%%%%%%%%%%%%%%%%%%%%%%
%%%%%%%%%%%%%%%%%%%%%%%%%%%%%%%%%%%%%%%%%%%%%%%
In this work we address a way to capture scalar field solutions on static spacetimes by using BPS formalism and relaxing the general covariance condition. We focus on configurations where the background geometry describes topological black holes and present both analytical and numerical solutions, in addition to discussing the use of conserved charges associated to such field configurations. The obtained solutions are radially stable and the zero-mode arising from the stability equation can be written analytically.
\end{abstract}

%\pacs{04.50.-h, 11.27.+d}
%%%%%%%%%%%%%%%%%%%%%%%%%%%%%%%%%%%%%%%%%%%%%%%
%%%%%%%%%%%%%%%%%%%%%%%%%%%%%%%%%%%%%%%%%%%%%%%
\maketitle
%%%%%%%%%%%%%%%%%%%%%%%%%%%%%%%%%%%%%%%%%%%%%%%
%%%%%%%%%%%%%%%%%%%%%%%%%%%%%%%%%%%%%%%%%%%%%%%
\section{Introduction}
%%%%%%%%%%%%%%%%%%%%%%%%%%%%%%%%%%%%%%%%%%%%%%%
%%%%%%%%%%%%%%%%%%%%%%%%%%%%%%%%%%%%%%%%%%%%%%%
The relationship between scalar fields and black holes has long been an object of discussion in the literature. In part, this is due to the various limitations imposed by no-hair theorems on the formation of black holes in scalar-tensor models (nice reviews on this subject are \cite{sotiriou2015black,herdeiro2015asymptotically}). They indicate that stable, asymptotically flat, spherically symmetric black holes arising from a large set of scalar-tensor theories must be essentially the same as those arising in General Relativity \cite{sotiriou2012black}. Recently, the interest in scalar solutions around black holes has been reheated due to scalarization phenomena, where a non-minimal coupling of the scalar field with curvature or matter sources is adjusted in such a way that a tachyonic instability arising from the field induces a phase transition in the strong gravity regime \cite{damour1993nonperturbative,silva2018spontaneous,herdeiro2018spontaneous,doneva2018new,blazquez2018radial,myung2019instability,fernandes2019spontaneous,brihaye2019scalarised,cunha2019spontaneously,brihaye2019spontaneous,astefanesei2020higher}.

In the standard cases black hole solutions have event horizons with spherical symmetry (see \cite{tangherlini1963schwarzschild,myers1986black} and references therein), but nowadays we know that black solutions can also be found with horizons presenting different  - planar or hyperbolic - topologies. In these cases they are called {\it topological black holes}. The study of topological black holes intensified in the 1990s mainly through extensions of the BTZ solution \cite{banados1992black} by using convenient coordinate identifications on AdS spacetime which generate solutions equipped with event horizons of different topologies \cite{aaminneborg1996making,brill1996multi,vanzo1997black}. Discussions about the possibility of forming these solutions in physical processes appear in \cite{mann1997pair,smith1997formation} and its generalizations for arbitrary dimensions first appeared in \cite{birmingham1999topological,cai1999topological}. In the following years, a better understanding of the qualitative differences of these solutions has been engendered in different ways (see \cite{brill1997thermodynamics,banados1998constant,cai1998topological,banados1998anti,emparan1999ads,aros2001black,martinez2004exact,martinez2006topological,birmingham2007stability,nadalini2008thermodynamical,stetsko2019topological,mahapatra2020exact} and related references).

Another way to approach models where scalar fields surround black holes occurs if one takes the effective limit where the background geometry is not sensitive to the presence of the scalar field. Discussions about the behavior of scalar solutions on fixed static backgrounds have arisen around extensions of the Derrick's Theorem \cite{derrick1964comments,hobart1963instability} for nonbackreacting curved spacetimes \cite{palmer1979derrick,radmore1978non,carloni2019derrick}, which set restrictions on the existence of stable scalar fields. In summary, these discussions indicates that stable spatially localized solutions arising from standard covariant descriptions of purely scalar relativistic Lagrangians can only exist in 1+1 dimensions in flat spaces and cannot exist on static asymptotically flat spacetimes in general. 

Outside the scope of these results some solutions have been found in models which do not satisfy the necessary conditions for the implementation of Derrick's Theorem and its extensions \cite{perivolaropoulos2018gravitational,alestas2019evading,morris2021radially,moreira2022analytical}. From prior knowledge on scalar field models in flat spacetime, we know that a way to evade Derrick's theorem in flat spacetimes emerge when the general covariance condition is relaxed allowing self-interacting scalar potentials to explicitly depends on background coordinates \cite{bazeia2003new,casana2015trapping}, which implies the violation of the momentum invariance due to the existence of preferred points in the model. This idea has been used to study field solutions in various models in planar setups (recent works on this subject are in \cite{bazeia2018dirac,bazeia2019configurational,casana2020bps,bazeia2021novel,bazeia2018magnetic,andrade2019first,bazeia2021configurational}) and has been extended to capture soliton-like structures in systems where the scalar field acts on static asymptotically flat four-dimensional setups \cite{morris2021radially}, in  $D$-dimensional geometries presenting anisotropic scaling properties \cite{moreira2022analytical} and in the modelling of dark matter scenarios \cite{correa2021modeling}.

In this work we bring the discussions about the existence of classical scalar field solutions on curved spacetimes for the generic case of static background geometries in effective models where the action is not completely covariant, but explicitly depends on the metric coordinates. We present a way to find localized kink-like scalars by using first-order equations (BPS formalism) and apply it on geometries denoting topological black holes, where the event horizon naturally enters the game as a preferred point. The role of conserved charges is discussed and we propose an alternative way of using them to describe solutions arising from the systems we are dealing with. The present approach extends the ideas presented in \cite{moreira2022analytical} and shows that it is possible to find well-behaved field solutions on fixed background geometries if we violate covariance and require that scalar self-interaction potential explicitly depends on the geometry. 

This work is organized as follows. In Sec. II we present the model setup, describing its scalar action, background geometry, field equations and boundary conditions. In Sec. III we develop the first-order formalism and discuss the role of conserved charges. In Sec. IV  and Sec. V we present analytical and numerical solutions on different black hole spacetimes, respectively. In Sec. VI we discuss the radial stability of the field solutions and in Sec. VII we close the work presenting some ending comments.
%%%%%%%%%%%%%%%%%%%%%%%%%%%%%%%%%%%%%%%%%%%%%%%
%%%%%%%%%%%%%%%%%%%%%%%%%%%%%%%%%%%%%%%%%%%%%%%
\section{General setup}
%%%%%%%%%%%%%%%%%%%%%%%%%%%%%%%%%%%%%%%%%%%%%%%
%%%%%%%%%%%%%%%%%%%%%%%%%%%%%%%%%%%%%%%%%%%%%%%
In this work we deal with classical scalar field models described by the $D$-dimensional action
\begin{equation}\label{action}
S_{\left(\phi\right)}=\int d^{D} x\sqrt{-g}\left(-\frac{1}{2}\nabla_a \phi\nabla^a\phi-V(x,\phi)\right),
\end{equation}
where $g=\text{det}(g_{ab})$ is the metric determinant and $\phi(x)$ denotes a scalar field which self-interacts through an scalar potential $V(x,\phi)$, which we require to explicitly depend on spacetime coordinates $x^a$, $a=0,1,2,\cdots, D-1$. The scalar field equation in this case is given by 
\begin{equation}\label{fieldeq}
\Box \phi=\frac{\partial V}{\partial\phi},
\end{equation}
where $\Box=g^{ab}\nabla_a\nabla_b$ is the d'Alembertian operator and the associated energy-momentum tensor is  
\begin{equation}\label{emt}
T_{ab}=\nabla_a\phi\nabla_b\phi-\frac{1}{2}g_{ab}\left(\nabla\phi\right)^2-g_{ab}V(x,\phi).
\end{equation}
We are interested in finding well-behaved spatially localized solutions on fixed background geometries generically expressed as
\begin{equation}\label{backmetric}
ds^2=-e^{2\nu(r)}dt^2+e^{2\lambda(r)}dr^2+ r^2\Hat{\sigma}_{ij}(x^k)dx^i dx^j,
\end{equation}
with $(x^0,x^1)\equiv (t,r)$, $i,j=2,3,\cdots,D-1$. Spacetimes described by the metric \eqref{backmetric} are static and exhibit radial symmetry on the $r$-coordinate, which is foliated by transverse sections describing constant curvature $(D-2)$-dimensional surfaces $\Hat{\Sigma}_{\gamma}$, with coordinates $x^i$ and metric $\Hat{\sigma}_{ij}(x^k)$. We restrict our study to cases where $\Hat{\Sigma}_{\gamma}$ represents closed Einstein manifolds with Ricci tensor
\begin{equation}
    \hat{R}_{ij}(\Hat{\sigma})=(D-3)\gamma\Hat{\sigma}_{ij}
\end{equation}
and curvature $\hat{R}=\Hat{\sigma}^{ij}\hat{R}_{ij}=(D-2)(D-3)\gamma$. In this way, it can be locally isometric to a sphere $S^{D-2}$ for positive $\gamma$, an Euclidean space $\mathbb{R}^{D-2}$ for $\gamma=0$ or an  hyperbolic space $H^{D-2}$ for negative $\gamma$ \cite{cai1998topological,brill1997thermodynamics}. Moreover, the background geometry has a time-like Killing vector $\xi=-\partial_t$, so one can associate a conserved current to the scalar field as $J^a=T^{ab}\xi_b$. Therefore, the energy of the scalar becomes
\begin{equation}\label{consch}
E(\xi)=-\int_{\Sigma }d^{D-1} x\sqrt{|h|} n_a\xi_b T^{ab},
\end{equation}
where $n_a=-e^{\nu}\delta_{a0}$ is the unit normal vector on the $(D-1)$-dimensional surface $\Sigma$ defined at fixed time and equipped with coordinates $x^p$ and induced metric $h_{pq}$ with $h=\text{det}(h_{pq})$ and $p,q=1,2\cdots,D-1$. We also consider cases where both scalar field and potential depends only on the radial coordinate, i.e.,
\begin{equation}
\phi=\phi (r) ~~\text{and}~~V(x,\phi)=V(r,\phi).
\end{equation}
Under these assumptions, the field equation \eqref{fieldeq} now is
\begin{equation}\label{2ordereq}
\frac{1}{r^{D-2}}e^{-\left(\nu+\lambda\right)}\frac{d~}{dr}\left(r^{D-2}e^{\nu-\lambda}\frac{d\phi}{dr}\right)=\frac{\partial V}{\partial\phi}
\end{equation}
and regularity on the solutions demands the boundary conditions
\begin{subequations}
\label{boundcond}
\begin{eqnarray}
   \phi(r\to r_0)&=&\phi_{0},~~\phi(r\to\infty)=\phi_{\infty},\\[4pt]
\lim_{r\to r_0}\left|\frac{d\phi}{dr}\right|&<&\infty,~~\lim_{r\to \infty}\frac{d\phi}{dr}=0,
\end{eqnarray}
\end{subequations}
where $\phi_0$ and $\phi_{\infty}$ are constants depending on the scalar field model we use and $r_0\geq0$ is the lower limit of the radial coordinate range, which depends on the geometry we use. 

Another point of interested we have is on finding conserved currents and charges which could help us to hold some analytical quantities even in situations where we do not have analytical solutions for the scalar field. We know from studies in $(1+1)$-dimensional flat spacetime that when dealing with kink-like solutions, stability is associated to the existence of the topological current $j^a=\epsilon^{ab}\partial_b\phi$ and its related charge is used to differ kink from anti-kink solutions within the topological sectors where the fields are placed on \cite{vilenkin1994cosmic}. In curved spacetimes this current keeps the same expression and leads to the conserved quantity
\begin{equation}\label{tc}
    Q=-\int_{\Sigma}d^{D-1}x\sqrt{|h|}\epsilon^{ab}n_a\partial_b\phi,
\end{equation}
but in this case its topological interpretation is lost and its  evaluation is highly dependent on the background geometry nonlinearities. One can evade this problem defining an alternative conserved current $\widetilde{J}^{a}$ based on a 1-form given by
\begin{equation}
\widetilde{A}=\widetilde{A}_a dx^a=e^{\nu+\lambda}\Delta\phi(r)\zeta(r)dt, 
\end{equation}
where $\Delta\phi(r)=\phi(r)-\phi_0$ and
\begin{eqnarray}\label{intpot}
   \zeta(r)=\left\{
\begin{array}{rcl}
&-\frac{1/(D-3)}{r^{D-3}}&, ~~\text{if}~~ D\neq 3,\\[4pt]
&\ln r&, ~~\text{if}~~ D=3.
\end{array}
\right.
\end{eqnarray}
We use the vector field above to build an auxiliary anti-symmetric tensor $\widetilde{f}_{ab}=\partial_a \widetilde{A}_b-\partial_b \widetilde{A}_a$ and relate it to the conserved current $\widetilde{J}^{a}=\nabla_b \widetilde{f}^{ab}$, which satisfies $\nabla_a \widetilde{J}^a=0$
and leads to the conserved charge arising from Gauss's law,
\begin{equation}
\label{q1f}
\widetilde{Q}=-\oint_{\partial\Sigma_\infty} d^{\small{D-2}}x \sqrt{|h^{(2)}|} n_a s_b \widetilde{f}^{ab},
\end{equation}
where $s^a=e^{-\lambda}\delta^a_1$ is the spacelike unit normal vector to the boundary of $\Sigma$ defined at fixed $r$ and calculated at spatial infinity - denoted here as $\partial\Sigma_\infty$ - and equipped with metric $h_{ij}^{(2)}=r^2\hat{\sigma}_{ij}$ with $h^{(2)}=\text{det}\left(h_{ij}^{(2)}\right)$. In particular, on the metric \eqref{backmetric} we have
\begin{subequations}
\begin{eqnarray}
Q&=&\omega_{D-2}^{(\gamma)}\int_{r_0}^{\infty}dr r^{D-2}e^{\nu+\lambda}\phi',\\[3pt]
\widetilde{Q}&=& \omega_{D-2}^{(\gamma)}\Delta\phi_{\infty} \lim_{r\to\infty}\left(1+r^{D-2}\left(\nu+\lambda\right)'\zeta(r)\right),~~~~~~~
\end{eqnarray}
\end{subequations}
where $\omega_{D-2}^{(\gamma)}=\oint_{\Hat{\Sigma}_\gamma}d^{D-2}x\sqrt{|\Hat{\sigma}_\gamma|}$
is the volume of the closed transverse space $\Hat{\Sigma}_{\gamma}$ and $\Delta\phi_{\infty}=\phi_{\infty}-\phi_0$. Note that the charges are engedered in such a way to coincides in 1+1 flat spacetime  $(Q=\widetilde{Q}=\Delta\phi_{\infty})$. Thus, the evaluation of the integral \eqref{q1f} is analytical in particular for asymptotically flat spacetimes and for spacetimes written in Schwarzschild coordinates $\left(\nu=-\lambda\right)$, which denote the structure of most existing black hole solutions in literature and where both charges $Q$ and $\widetilde{Q}$ are completely determined by the boundary values of the scalar field only. In this way one can replace the charge $Q$ by the new charge $\widetilde{Q}$ in the characterization of scalar solutions. 
%%%%%%%%%%%%%%%%%%%%%%%%%%%%%%%%%%%%%%%%%%%%%%%
%%%%%%%%%%%%%%%%%%%%%%%%%%%%%%%%%%%%%%%%%%%%%%%
\section{BPS formalism}
%%%%%%%%%%%%%%%%%%%%%%%%%%%%%%%%%%%%%%%%%%%%%%%
%%%%%%%%%%%%%%%%%%%%%%%%%%%%%%%%%%%%%%%%%%%%%%%
Even with all the simplifications we made so far, hard nonlinearities could still occur in the second-order equation \eqref{2ordereq} which would make its treatment quite complicated. To overcome some of these difficulties, we implement the BPS (or First-order) formalism in our study, which provides a simpler path to find minimal energy solutions by using first-order equations. In order to set the necessary routine we point out that within the integrand of \eqref{consch} we have $-\eta_a \xi_b T^{ab}=-e^{\nu}T^0_{~0}$ 
and that the $00$-component of the energy-momentum tensor can be expressed as
\begin{equation} \label{too}
-T^0_{~0}=\frac{1}{2}\left(e^{-\lambda}\frac{d\phi}{dr}\mp\sqrt{2V}\right)^{2}\pm e^{-\lambda}\frac{d\phi}{dr}\sqrt{2V}.
\end{equation}
Therefore, the energy of the scalar field in \eqref{consch} has a lower bound expressed by the inequality
\begin{equation}\label{bpsineq}
    E(\xi)\geq\pm\int_{\Sigma} d^{D-1} x\sqrt{|h|}e^{\nu-\lambda}\frac{d\phi}{dr}\sqrt{2V},
\end{equation}
which is saturated for cases where
\begin{equation}\label{1oeq}
\frac{d\phi}{dr}=\pm e^\lambda\sqrt{2V}.
\end{equation}
The above equation is first order and therefore simpler to handle than the original second order field equation \eqref{2ordereq}. Moreover, a direct calculation shows that fields satisfying equation \eqref{1oeq} also solves the field equation \eqref{2ordereq}. By construction, solutions whose field equation saturates the inequality \eqref{bpsineq} have minimal energy and in this way we call them {\it BPS solutions} \cite{prasad1975exact,bogomol1976stability}, where its BPS energy is defined as $E_{BPS}=\text{min}\{E(\xi)\}$. BPS solutions naturally satisfies the Weak Energy Condition 
\begin{equation}\label{wec}
\rho=e^{-2\lambda}\left(\frac{d\phi}{dr}\right)^2=T_{ab}\xi^a\xi^b\geq0,
\end{equation}
where $\xi_a\xi^a=-1$, which implies that these fields are well behaved in the sense that its energy density measured by observers on time-like curves is always nonnegative \cite{kontou2020energy}. Explicitly written in terms of the field, the BPS energy is given by
\begin{equation}\label{bpsdef}
    E_{BPS}=\omega_{D-2}^{(\gamma)}\int_{r_0}^\infty dr r^{D-2} e^{\nu-\lambda}\left(\frac{d\phi}{dr}\right)^2.
\end{equation}
Since all finite quantities are proportional to the volume term $\omega_{D-2}^{(\gamma)}$, one must be careful when defining the surface $\hat{\Sigma}_{\gamma}$ to ensure that it is closed. This requirement is naturally satisfied if $\hat{\Sigma}_{\gamma}$ has a spherical topology and can be implemented by performing an identification $x^i\to x^i+l^1$ in cases where $\hat{\Sigma}_{\gamma}$ is planar. For cases where $\hat{\Sigma}_{\gamma}$ has negative curvature its topology is $H^{D-2}/\Gamma$, where $\Gamma$ denotes an specific discrete subgroup of $H^{D-2}$ \cite{vanzo1997black,brill1997thermodynamics,birmingham2007stability,cai1998topological,martinez2006topological}.

A direct way to set a step-by-step method to capture minimal energy scalar field solutions is defining an auxiliary function $W(\phi)$ such that
\begin{equation}\label{1ordereq}
\frac{d\phi}{dr}=\pm \frac{e^{\lambda-\nu}}{r^{D-2}}\frac{dW}{d\phi}.
\end{equation}
Under this choice the scalar potential becomes
\begin{equation}\label{potmodel}
V(r,\phi)=\frac{1}{2}\frac{e^{-2\nu}}{r^{2\left(D-2\right)}}\left(\frac{dW}{d\phi}\right)^2,
\end{equation} 
 which explicitly depends on the $r$-coordinate, as required, and have all terms holding the scalar field encoded within the function $W(\phi)$. An advantage of this formalism is that even if one doesn't know the exact solution of the scalar field it is immediate to find its energy, now given by
\begin{equation}\label{ebps}
    E_{BPS}=\omega_{D-2}^{(\gamma)}|\Delta W|,
\end{equation}
where $\Delta W=W(\phi_\infty)-W(\phi_0)$. The BPS energy is easily determined once we know $W(\phi)$ and the pair $(\phi_0,\phi_\infty)$. 

The structure of the scalar potential \eqref{potmodel} and the first-order equation \eqref{1ordereq} indicates that field solutions must appear in pairs, one for each sign in  equation \eqref{1ordereq}. The boundary conditions \eqref{boundcond} indicate that the field must interpolate between distinct values if we want to find nonzero conserved charges. Regularity at the boundaries requires $\phi_0$ and/or $\phi_\infty$ to be extrema of $W(\phi)$ in such a way that the scalar potential is set to zero there. Moreover, since the derivative of the scalar field doesn't change its signal, we can infer that the solutions we are looking for must have a kink-like profile.

The machinery described so far works both for asymptotically flat spacetimes and for geometries that can be expressed in Schwarzschild coordinates, which is the case for most known black hole solutions. Therefore, we particularize our applications on black holes of different dimensions and topologies. In all examples we present below we use as auxiliary function
\begin{equation}
    W(\phi)=\phi-\frac13\, {\phi^3}.
\end{equation}
 Since we are interested in describe scalar fields surrounding black hole solutions, we also require that the scalar field approaches zero as it moves away from the event horizon $r_0=r_h$, so we set $\phi_\infty=0$. Moreover, in order to evade divergences at the horizon, we set $\phi_0$ as a minimum of $W(\phi)$. Consequently $\left(\phi_0,\phi_\infty\right)=(\pm1,0)$ for any case ahead. 
%%%%%%%%%%%%%%%%%%%%%%%%%%%%%%%%%%%%%%%%%%%%%%%
%%%%%%%%%%%%%%%%%%%%%%%%%%%%%%%%%%%%%%%%%%%%%%%
\section{Analytical solutions}
%%%%%%%%%%%%%%%%%%%%%%%%%%%%%%%%%%%%%%%%%%%%%%%
%%%%%%%%%%%%%%%%%%%%%%%%%%%%%%%%%%%%%%%%%%%%%%%
\subsection{BTZ black hole}
%%%%%%%%%%%%%%%%%%%%%%%%%%%%%%%%%%%%%%%%%%%%%%%
%%%%%%%%%%%%%%%%%%%%%%%%%%%%%%%%%%%%%%%%%%%%%%%
%---------------------------------------------%
\begin{figure}[t!]	
		\centering
\includegraphics[width=8cm]{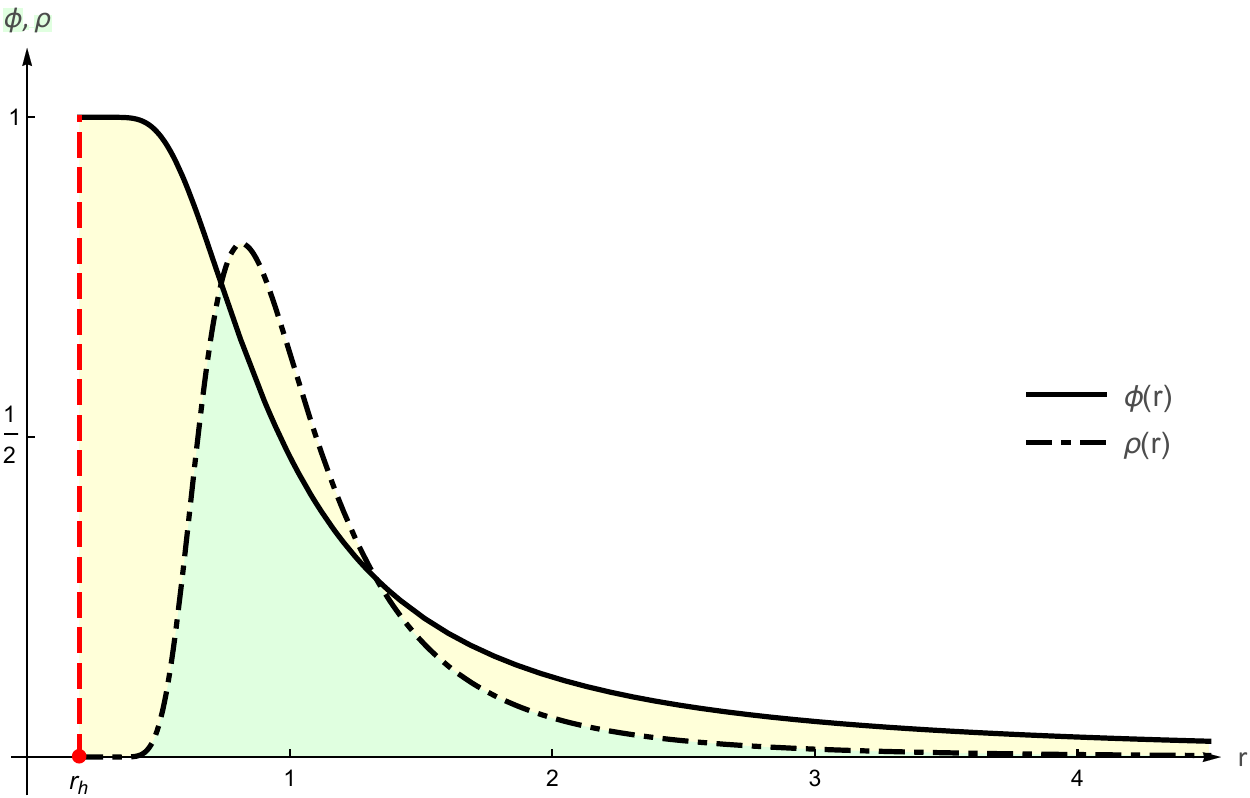}\\
\vspace{2mm}
\caption{Scalar field profile (anti-kink solution) and its energy density on the BTZ black hole with $M=0.005$ and $r_h=\sqrt{8M}$. }
\label{fig1}
\end{figure}
%---------------------------------------------%
The first case we explore which allows analytical treatment is the BTZ solution \cite{banados1992black}, which describes a topological black hole in 2+1 dimensions whose event horizon presents a negative curvature. Its nonrotating setup has background metric given by
\begin{equation}
    ds^2=-\left(-8M+r^2\right)dt^2+\frac{dr^2}{-8M+r^2}+r^2d\varphi^2,
\end{equation}
where $M$ denotes the black hole mass, the event horizon occurs at $r_h=\sqrt{8M}$ and there exists a coordinate identification $\varphi\sim\varphi+2\pi$. Thus, the scalar potential is
\begin{equation}
    V(r,\phi)=\frac{\left(1-\phi^2\right)^2}{2r^2\left(r^2-8M\right)}
\end{equation}
and the first-order equation which determines the scalar field solution becomes
\begin{equation}
\frac{d\phi}{dr}=\pm\frac{1-\phi^2}{r(r^2-8M)}.
\end{equation}
The equation above is satisfied by the solution 
\begin{equation}
    \phi(r)=\pm\tanh\left(\frac{1}{16M}\ln\left(1-\frac{8M}{r^2}\right)\right),
\end{equation}
 and in this way we clearly have $\left(\phi_0,\phi_\infty\right)=(\mp1,0)$, as expected and so the BPS energy becomes
\begin{equation}
    E_{BPS}=\frac43 \pi,
\end{equation}
which is independent of the black hole mass. The energy density of the scalar field  is given by equation \eqref{wec} and evaluated as
 \begin{equation}
     \rho(r)=\frac{\text{sech}^4\left(\frac{1}{16 M}\log \left(1-\frac{8 M}{r^2}\right)\right)}{r^2\left(r^2-8 M\right)}
 \end{equation}
and non-negative for $r\geq r_h$. The profile of the scalar field and the energy density is depicted in Fig. (\ref{fig1}). Note that both the field and the energy density are regular and divergence-free outside the event horizon, where the solution is valid. 

For pedagogical reason, let us now investigate both charges \eqref{tc} and \eqref{q1f} to see how they differ from each other. The first conserved charge is given by the integral
\begin{equation}
    Q=\pm 2\pi\int_{\sqrt{8M}}^{\infty}dr \frac{\text{sech}^2\left(\frac{1}{16 M}\ln \left(1-\frac{8 M}{r^2}\right)\right)}{r^2-8 M}<\infty,
\end{equation}
which is convergent, but not solvable in general. The second charge is simpler, leading to 
\begin{equation}
    \widetilde{Q}= \pm 2\pi,
\end{equation}
which is also independent of the black hole parameters and where the positive (negative) sign denotes the kink (anti-kink) solution.
%%%%%%%%%%%%%%%%%%%%%%%%%%%%%%%%%%%%%%%%%%%%%%%
%%%%%%%%%%%%%%%%%%%%%%%%%%%%%%%%%%%%%%%%%%%%%%%
\subsection{Schwarzschild-Tangherlini black hole}
%%%%%%%%%%%%%%%%%%%%%%%%%%%%%%%%%%%%%%%%%%%%%%%
%%%%%%%%%%%%%%%%%%%%%%%%%%%%%%%%%%%%%%%%%%%%%%%
%---------------------------------------------%
\begin{figure}[t!]	
		\centering
\includegraphics[width=8cm]{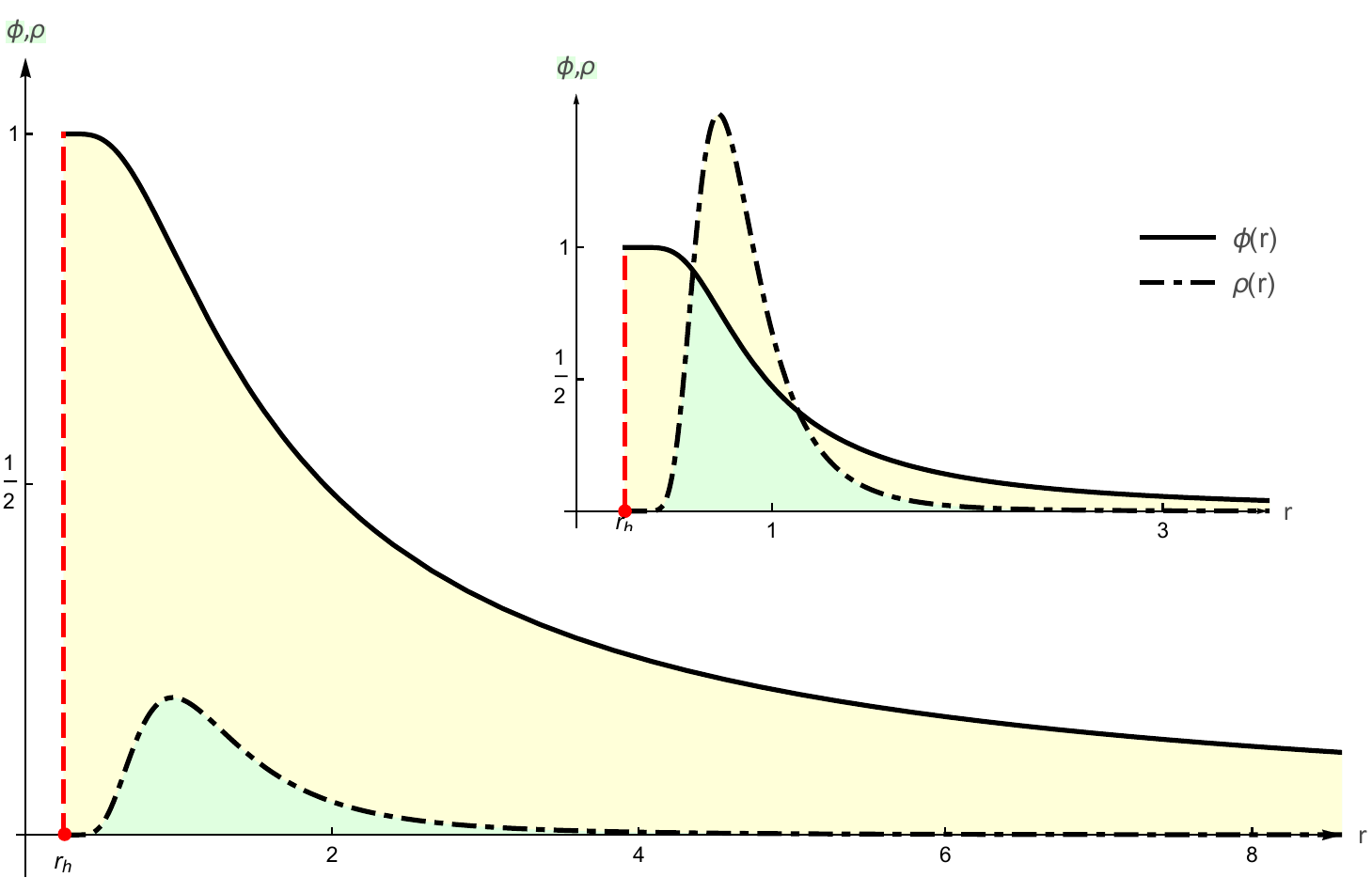}\\
\vspace{2mm}
\caption{Scalar field profile (anti-kink solution) and its respective energy density on the Schwarzschild-Tangherlini black hole for $D=3+1$ and $D=4+1$ (inset), with $r_h=1/4$ and $m=r_h^{D-3}/2$ in both cases. }
\label{fig2}
\end{figure}
%---------------------------------------------%
We continue by presenting analytical solutions on Schwarzschild-Tangherlini black holes \cite{tangherlini1963schwarzschild}, which generalizes the Schwarzschild solution to higher dimensions. This solution presents an event horizon with spherical topology $(\gamma=1)$. For $D>3$ we have
\begin{equation}
    ds^2=-e^{2\nu(r)}dt^2+\frac{dr^2}{e^{2\nu(r)}}+r^2d\Omega_{D-2}^2,~~\text{where}
~~e^{2\nu(r)}=e^{-2\lambda(r)} =1-\frac{2m}{r^{D-3}} 
\end{equation}
 and $d\Omega_{D-2}^2$ represents the unit element volume of a $(D-2)$-dimensional sphere. The $m$-parameter is related to the black hole mass as $M=(D-2)\omega_{D-2}^{(1)}m/8\pi$.
On this geometry the scalar field potential becomes
\begin{equation}
    V(r,\phi)=\frac{\left(1-\phi^2\right)^2}{2r^{2(D-2)}\left(1-\frac{2m}{r^{D-3}}\right)}
\end{equation}
and thus we have to deal with the scalar field equation 
\begin{equation}\label{foeq2}
    \frac{d\phi}{dr}=\pm\frac{1}{r^{D-2}}\frac{1-\phi^2}{\left(1-\frac{2m}{r^{D-3}}\right)},
\end{equation}
which is solved by
\begin{equation}\label{sch}
    \phi(r)=\pm\tanh\left(\frac{1/2m}{(D-3)}\ln\left(1-\frac{2m}{r^{D-3}}\right)\right).
\end{equation}
The above solution is regular on the horizon situated at $r_h=\left(2m\right)^{1/(D-3)}$ for any $D>3$ and, as in the previous case, we again have $\left(\phi_0,\phi_\infty\right)=(\mp1,0)$. The energy density is now given by
\begin{equation}
    \rho(r)=\frac{ \text{sech}^4\left(\frac{1/2m}{(D-3)}\ln\left(1-\frac{2m}{r^{D-3}}\right)\right)}{r^{2(D-2)}\left(1-\frac{2m}{r^{D-3}}\right)},
\end{equation}
which is depicted along with the scalar field in Fig. \eqref{fig2}. The BPS energy and the charge associated to the field solution in this case are
\begin{equation}
    E_{BPS}=\frac43\frac{\pi^{\frac{D-1}{2}}}{\Gamma\left(\frac{D-1}{2}\right)},
\end{equation}
and $\widetilde{Q}=\pm(3/2)E_{BPS}$; again, they are independent of the black hole mass. It is worth mentioning that the particular case of the solution \eqref{sch} for $D=3+1$ was previously obtained in \cite{morris2021radially} by a different method. Furthermore, the asymptotic behavior of the scalar field is
\begin{equation}
    \phi(r\to\infty)\simeq-\frac{3}{2}\frac{E_{BPS}/\widetilde{Q}}{r}+\mathcal{O}\left(\frac{1}{r^2}\right),
\end{equation}
so we can identify the scalar charge arising in the leading term of the asymptotic expansion as $q_s=-3E_{BPS}/2\widetilde{Q}$, which may indicate that the nature of these charges, which are important in the scalarization process, may have information about the energy of the field along with some underlying conservation law.
%%%%%%%%%%%%%%%%%%%%%%%%%%%%%%%%%%%%%%%%%%%%%%%
%%%%%%%%%%%%%%%%%%%%%%%%%%%%%%%%%%%%%%%%%%%%%%%
\section{Numerical Solutions}
%%%%%%%%%%%%%%%%%%%%%%%%%%%%%%%%%%%%%%%%%%%%%%%
%%%%%%%%%%%%%%%%%%%%%%%%%%%%%%%%%%%%%%%%%%%%%%%
For $D>3$ we cannot analytically deal with charged black holes equipped with a cosmological constant within the formalism we developed above. Therefore, now it is necessary to implement a numerical approach in order to find field solutions on these geometries. We consider fields on topological black holes as those considered in Refs. \cite{cai1999topological,wu2003topological}, given by 
\begin{equation}\label{tbh}
    ds^2=-e^{2\nu(r)}dt^2+\frac{dr^2}{e^{2\nu(r)}}+r^2\Hat{\sigma}_{ij}dx^i dx^j,~~\text{where}~~
   e^{2\nu(r)}=e^{-2\lambda(r)} =\gamma-\frac{2m}{r^{D-3}}+\frac{q^2}{r^{2(D-3)}}+\frac{r^2}{l^2}.
\end{equation}
The solution above arises as generalizations of the non-rotating BTZ solution \cite{banados1992black} for higher dimensions with
\begin{eqnarray}\label{intpotbtz}
   \Hat{\sigma}_{ij}dx^i dx^j=\left\{
\begin{array}{rcl}
d\Omega_{D-2}^2, ~&\text{if}&~ \gamma=1,\\[2pt]
\smashoperator[r]{\sum_{j=2}^{D-1}}dx^j dx^j, ~&\text{if}&~ \gamma=0,\\[9pt]
dH_{D-2}^2, ~&\text{if}&~ \gamma=-1,
\end{array}
\right.
\end{eqnarray}
where  $d\Omega_{D-2}^2$ and $dH_{D-2}^2$ represents the unit metric on $S^{D-2}$ and $H^{D-2}$, respectively. These backgrounds asymptotically approaches $AdS$ spacetimes with cosmological constant $\Lambda=-(D-1)(D-2)/2l^2$. In addition, the black hole mass and charge are given by
\begin{equation}
M=\frac{
(D-2)\omega_{D-2}^{(\gamma)}}{8\pi}m~~\text{and}~~Q=\pm\frac{\omega_{D-2}^{(\gamma)}}{4\pi}q,
\end{equation}
  implicitly related to event horizons as
\begin{equation}
    q^2=r_{h}^{D-3}\left(2m-r_h^{D-3}\left(\gamma+\frac{r_h^2}{l^2}\right)\right).
\end{equation}
Since we are not looking directly at the models which produce the background geometries we are dealing with, we assume that the charged Einstein-Tangherlini solution arises here at the limit $(\Lambda,\gamma)\to (0,1)$. 

\begin{figure}	
		\centering
\includegraphics[width=8cm]{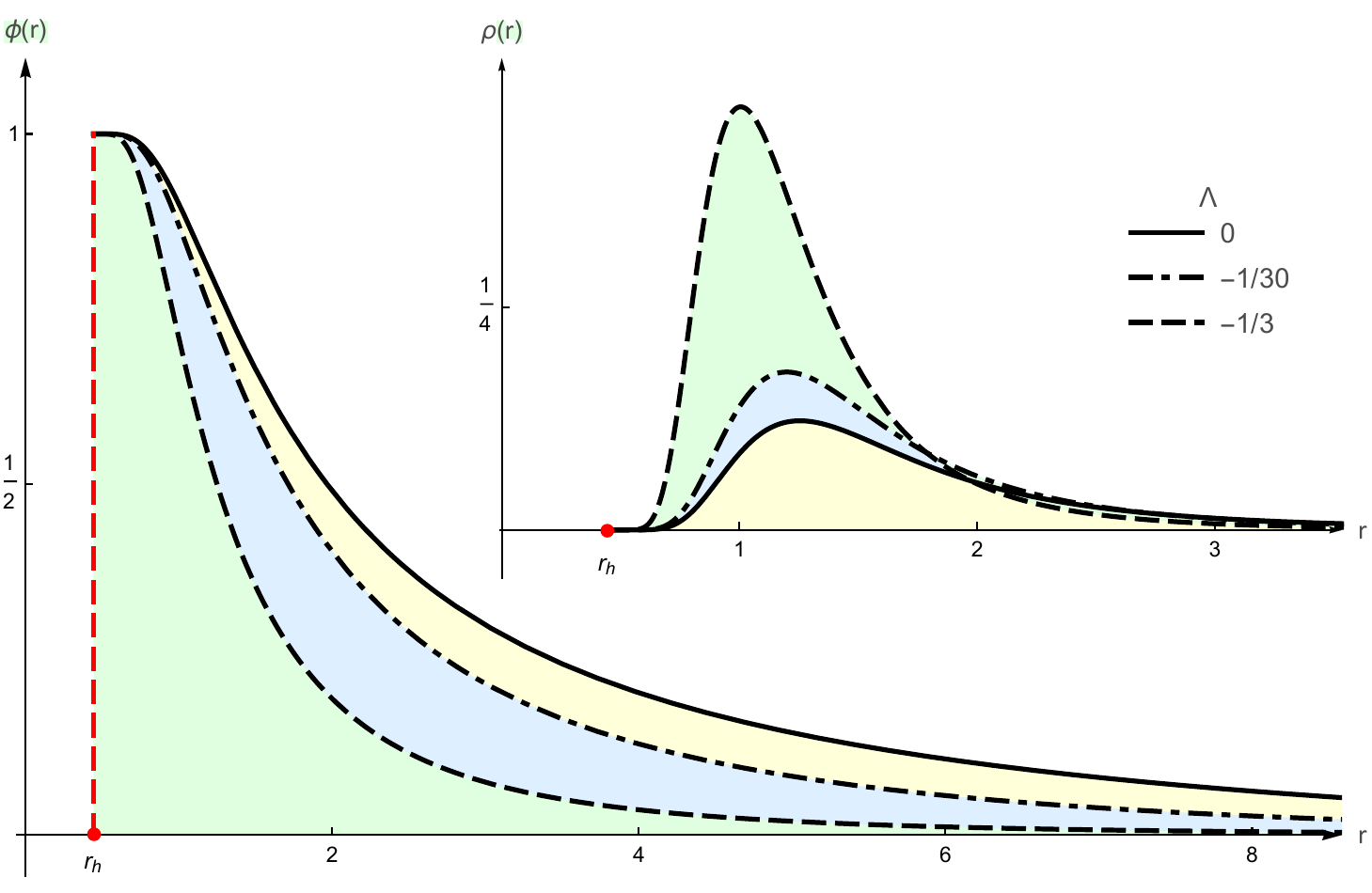}\\
\vspace{2mm}
\caption{Scalar field profile (anti-kink solution) and energy density (inset) on the topological black hole in $D=3+1$ with $r_h=1/4,~|q|=1/3,~\gamma=1$ and $m=\left(1-3\Lambda/100\right)m_0$, where $m_0=25/72$, for some values of $\Lambda$.  }
\label{fig3}
\end{figure}

According to all these considerations, the scalar potential guiding the field self-interactions is
\begin{equation}
    V(r,\phi)=\frac{\left(1-\phi^2\right)^2}{2r^{2(D-2)}\left(\gamma-\frac{2m}{r^{D-3}}+\frac{q^2}{r^{2(D-3)}}+\frac{r^2}{l^2}\right)}
\end{equation}
and the field solution can be captured by the first-order field equation 
\begin{equation}\label{eqtbh}
    \frac{d\phi}{dr}=\pm\frac{
    1-\phi^2}{r^{D-2}\left(\gamma-\frac{2m}{r^{D-3}}+\frac{q^2}{r^{2(D-3)}}+\frac{r^2}{l^2}\right)}.
\end{equation}
It is not solvable analytically, in general, but one can treat it numerically. The profile of the scalar field for some values of $\Lambda$ and its respective energy densities for $D=3+1$, $r_h=1/4,~|q|=1/3,~\gamma=1$ and $m=\left(1-3\Lambda/100\right)m_0$, where $m_0=25/72$ is the ``mass parameter" value for $\Lambda=0$, are depicted in Fig. \eqref{fig3}. 

%---------------------------------------------%
%---------------------------------------------%

For asymptotically $AdS_{(D)}$ spacetimes the field solutions have a kink-like shape since the cosmological constant acts as a field suppression term and forces the energy density to be more concentrated near the horizon as $\left|\Lambda\right|$ increases, as also indicated in Fig. \eqref{fig3}. The asymptotic analysis of \eqref{eqtbh} gives
\begin{equation}
    \phi\left(r\to\infty\right)\sim\pm\frac{(D-2)/2\Lambda}{r^{D-1}}+\mathcal{O}\left(\frac{\gamma}{r^{D+1}}\right)+\cdots,
\end{equation}
and reveals that here the existence of the cosmological constant ensures the field to satisfy the boundary conditions \eqref{boundcond}. For asymptotically flat spacetimes the behavior of the scalar depends on the curvature at the horizon. For $\Lambda=0$ and $\gamma=\pm1$ one finds well-behaved field solutions approaching infinity as 
\begin{equation}
    \phi\left(r\to\infty\right)\sim\mp\frac{1}{\gamma}\frac{1/(D-3)}{r^{D-3}}+\mathcal{O}\left(\frac{m}{r^{2(D-3)}}\right)+\cdots,
\end{equation}
so we can infer that in the absence of the cosmological constant the nonzero horizon curvature asymptotically acts on the leading term to ensure the kink-like profile of the scalar field. For $\Lambda=0$ and $\gamma=0$, the scalar asymptotically holds a logarithmic divergence as $\phi\left(r\to\infty\right)\sim \mp\frac{1}{2m}\ln (r)$, and in this way it doesn't have spatially localized structure. 

Even dealing with models which we were not able to get analytical field solutions, we can calculate its conserved charge and BPS energy  exactly for well-behaved cases where $\Lambda<0$ or $\left(\Lambda,\gamma\right)=\left(0,\pm1\right)$. The evaluation gives
\begin{equation}
    E_{BPS}=\frac23\; \omega_{D-2}^{(\gamma)},
\end{equation}
and $\widetilde{Q}=\pm(3/2)E_{BPS}$ which, again, are independent of the black hole parameters. However, for cases where $\left(\Lambda,\gamma\right)=\left(0,0\right)$,  neither $E_{BPS}$ nor $\widetilde{Q}$ are finite quantities since we don't have scalar solutions satisfying the full set of boundary conditions \eqref{boundcond}.
%%%%%%%%%%%%%%%%%%%%%%%%%%%%%%%%%%%%%%%%%%%%%%%
%%%%%%%%%%%%%%%%%%%%%%%%%%%%%%%%%%%%%%%%%%%%%%%
\section{Stability}
%%%%%%%%%%%%%%%%%%%%%%%%%%%%%%%%%%%%%%%%%%%%%%%
%%%%%%%%%%%%%%%%%%%%%%%%%%%%%%%%%%%%%%%%%%%%%%%
The stability analysis of the solutions found above is made from
the imposition of radial and time-periodic perturbations around static solutions such as $\phi(r,t)=\phi(r)+e^{i\omega t}\,\psi(r)$, which lead us to the equation
\begin{equation}\label{steq}
\left(-\Box+\left.\frac{\partial^2 V}{\partial\phi^2}\right|_{\phi=\phi(r)}\right)\psi(r)=\omega^2 e^{-2\nu}\psi(r),
\end{equation}
which can be rewritten as an Sturm-Liouville problem, 
\begin{equation}\label{stmlvl}
    \left(-\frac{d~}{dr}\left(p(r)\frac{d~}{dr}\right)+q(r)\right)\psi=\omega^2r^{D-2}e^{-(\nu-\lambda)}\psi,
\end{equation}
where
\begin{equation}
p(r)=r^{D-2}e^{\nu-\lambda}~~\text{and}~~
q(r)=r^{D-2}e^{\nu+\lambda}\left.\frac{\partial^2V}{\partial\phi^2}\right|_{\phi(r)}\!\!\!\!.
\end{equation}
In order to have stable setups, we must ensure that the fluctuations around the static solution are regular everywhere. In this sense, we must show that the frequency $\omega$ is real for the above Sturm-Liouville problem, since otherwise the field fluctuations exponentially diverges in time. One can infer from the properties of the Sturm-Liouville problem that the stability equation \eqref{steq} in general provides an infinite discrete tower of real bound states, as also occurs when considering kink-like models with generalized dynamics in flat spacetimes \cite{igor}. However, this property does not guarantee that there are no eigenstates with eigenvalues $\omega^2<0\in\mathbb{R}$, which indicate imaginary frequencies. In order to show that such states cannot have negative eigenvalues, first note that the inner product compatible with the stability equation \eqref{stmlvl} is \begin{equation}\label{inner}
    \langle \psi_m,\psi_n\rangle=\int_{r_o}^{\infty}dr r^{D-2}e^{-\left(\nu(r)-\lambda(r)\right)}\bar{\psi}_m(r)\psi_n(r),
\end{equation}
where $m,n\in \mathbb{N}$. After some manipulations one can show that equation \eqref{stmlvl} can be  factorized as 
$S^{\dagger}S\widetilde{\psi}=\omega^2\widetilde{\psi}$ where $\widetilde{\psi}=e^{\nu-\lambda}\psi$ and 
\begin{equation}
S^{\dagger}=e^{\nu-\lambda}\left(\frac{d~}{dr}+\mathcal{W}(r)+\alpha(r)\right),~~~
S~=e^{\nu-\lambda}\left(-\frac{d~}{dr}+\mathcal{W}(r)\right),
\end{equation}
with auxiliary functions 
\begin{equation}
\mathcal{W}(r)=\frac{W_{\phi\phi}}{r^{D-2}}e^{-(\nu-\lambda)}+\left(\nu-\lambda\right)',~~~
\alpha(r)=2\left(\nu-\lambda\right)'-\frac{D-2}{r}.
\end{equation}
By using results from \cite{hounkonnou2004factorization}, we conclude that the operators $S^{\dagger}$ and $S$ are mutually adjoint under the inner product \eqref{inner} along with the boundary condition $\left.r^{D-2}\psi_m(r)\psi_n(r)\right|_{r_0}^{\infty}=0$, which implies that the Hamiltonian $\hat{H}=S^\dagger S$ is non-negative. It ensures stability of the solutions found since it shows that equation \eqref{steq} does not allow states with negative energy, which implies that $\omega_n^2\geq 0$ for all $n\in \mathbb{N}$. In particular, the ground state is given by the zero mode evaluated at $S\widetilde{\psi}_0=0$, resulting in
\begin{equation}\label{zmode}    \psi_0=c\exp{\left(\int dr \frac{W_{\phi\phi}}{r^{D-2}}e^{-(\nu-\lambda)}\right)},
\end{equation}
where $c$ is a normalization constant.
%%%%%%%%%%%%%%%%%%%%%%%%%%%%%%%%%%%%%%%%%%%%%%%
%%%%%%%%%%%%%%%%%%%%%%%%%%%%%%%%%%%%%%%%%%%%%%%
\section{Ending comments} 
%%%%%%%%%%%%%%%%%%%%%%%%%%%%%%%%%%%%%%%%%%%%%%%
%%%%%%%%%%%%%%%%%%%%%%%%%%%%%%%%%%%%%%%%%%%%%%%
In this work we addressed the problem of obtaining spatially localized scalar field structures on static spacetimes. By implementing the BPS formalism, we were able to find minimal energy solutions once we relaxed the general covariance by allowing the scalar self-interaction potential to explicitly depend on metric coordinates. We applied the described routine in the search for kink-like fields on different topological black holes, presenting both analytical and numerical solutions. The solutions we found are regular, radially stable and divergence-free outside the event horizon. For topological black holes, the behavior of the scalar field results from suppression terms arising from non-zero curvature effects at the event horizon and at infinity. Thus, not all types of topological black holes support spatially localized solutions. In particular, we could not obtain such solutions on black holes with planar horizons in asymptotically flat spacetimes.

Due to difficulties in calculations when dealing with nonlinearities arising from curvature effects, it is not always practical to use the charge \eqref{tc} to characterize the field solution as a kink or anti-kink since it leads to integrals which are not simple to deal with in the presence of generic volume elements. In order to circumvent this problem, we have built an alternative charge \eqref{q1f} from a scalar field dependent 1-form, which reproduces the same value as the topological charge in $(1+1)$-dimensional flat spacetime and whose calculation is feasible from Gauss' Law for a wide variety of static geometries. 

Very recently, another system which also handles scalar fields on curved static spacetimes along with a general covariance violation was introduced in \cite{morris2022bps}. In this case, the scalar action does not have a self-interacting potential, but instead it has a Maxwell term coupled to a function explicitly dependent on the background geometry and interpreted as permittivity of the medium, which is used to induce the necessary instability for the scalarization processes. This suggests that we can consider extensions of the model investigated above inserting the Maxwell term coupled to the scalar through electrical permittivity in order to understand how the gauge field may affect the shape and stability of the solutions. Another interesting problem would be to study permittivity-equipped electric dipole models like the one discussed in \cite{bmm22} on curved background geometries to observe how the dipole affects the system. We hope to report on this in the near future.

%%%%%%%%%%%%%%%%%%%%%%%%%%%%%%%%%%%%%%%%%%%%%%%
%%%%%%%%%%%%%%%%%%%%%%%%%%%%%%%%%%%%%%%%%%%%%%%
\begin{acknowledgments}
%%%%%%%%%%%%%%%%%%%%%%%%%%%%%%%%%%%%%%%%%%%%%%%
%%%%%%%%%%%%%%%%%%%%%%%%%%%%%%%%%%%%%%%%%%%%%%%
DCM would like to thank the Brazilian agencies CNPq and FAPESQ-PB for the financial support (PDCTR FAPESQ-PB/CNPq, Grant no. 317985/2021-3). FAB and DB acknowledge support from CNPq (Grant nos. 312104/2018-9, 439027/2018-7 and 303469/2019-6, 404913/2018-0) and also CNPq/PRONEX/FAPESQ-PB (Grant nos. 165/2018 and 015/2019), for partial financial support.

%\appendix*\section{section name}

\end{acknowledgments}
%%%%%%%%%%%%%%%%%%%%%%%%%%%%%%%%%%%%%%%%%%%%%%%%%%%%%%%%%%
%%%%%%%%%%%%%%%%%%%%%%%%%%%%%%%%%%%%%%%%%%%%%%%%%%%%%%%%%%
\bibliography{biblio} 

%%%%%%%%%%%%%%%%%%%%%%%%%%%%%%%%%%%%%%%%%%%%%%%%%%%%%%%%%%
%%%%%%%%%%%%%%%%%%%%%%%%%%%%%%%%%%%%%%%%%%%%%%%%%%%%%%%%%%
\end{document}